\documentclass[a4paper,10pt,twoside]{cpc-hepnp}
\usepackage{hyperref}
\usepackage{multicol}
\usepackage{ulem}
\usepackage{graphicx}
\usepackage{booktabs}
\usepackage{amssymb,bm,mathrsfs,bbm,amscd}
\usepackage[tbtags]{amsmath}
\usepackage{lastpage}
\usepackage{rotating}
\usepackage{color,soul}
\setulcolor{blue}
\begin{document}



\title{Derived Born cross sections of $e^+e^-$ annihilation
into open charm mesons from CLEO-c measurements\thanks{Supported 
in part by National Natural Science
Foundation of China (NSFC) under contract Nos. 11235011,
11475187, 11521505, and U1632106; the Ministry of Science and Technology of
China under Contract No. 2015CB856701; Key Research Program of Frontier Sciences,
CAS, Grant No. QYZDJ-SSW-SLH011; and the CAS Center for Excellence
in Particle Physics (CCEPP).}}

\author{%
      Xiang-Kun Dong$^{1;1)}$\email{dongxiangkun14@mails.ucas.edu.cn}%
\quad Liang-Liang Wang$^{2;2)}$\email{llwang@ihep.ac.cn}%
\quad Chang-Zheng Yuan$^{2,1;3}$\email{yuancz@ihep.ac.cn}%
}
\maketitle

\address{%
$^1$ University of Chinese Academy of Sciences, Beijing 100049, China\\
$^2$ Institute of High Energy Physics, Chinese Academy of Sciences, Beijing 100049, China 
}

\begin{abstract}

The exclusive Born cross sections of the production of $D^0$,
$D^+$ and $D_s^+$ mesons in $e^+e^-$ annihilation at 13 energy
points between 3.970 and 4.260 GeV are obtained by applying
corrections for initial state radiation and vacuum polarization to
the observed cross sections measured by CLEO-c experiment. Both
the statistical and the systematic uncertainties for the obtained
Born cross sections are properly estimated.

\end{abstract}

\begin{keyword}
electron-positron annihilation, charm production, Born cross
section, radiative correction
\end{keyword}

\begin{pacs}
13.66.Bc, 13.25.Gv, 14.40.Rt
\end{pacs}


\begin{multicols}{2}

\section{Introduction}

In the energy range between the mass of $\psi(3770)$ (just above the smallest open charm production threshold 
of $D^0\bar{D^0}$) and 4.7 GeV
which contains quite a few open charm production thresholds, 
several vector charmonium-like structures (known as $Y$ states, e.g. $Y(4260)$, $Y(4360)$, $Y(4660)$ ...) 
were discovered in the $e^+e^-$ annihilation 
into the hidden-charm final states ~\cite{Yuan:2007sj,Ablikim:2014qwy,Aubert:2005rm,Aubert:2007zz,Wang:2007ea,Pakhlova:2008vn} 
over the past decade. The strong coupling of them to the hidden-charm final states and 
the difficulty in fitting all of the $Y$ states to the potential model prediction~\cite{potentialModle}
indicate that they could be exotic particle candidates, like hybrids, tetraquarks, meson molecules
and so on~\cite{XYZ-review2016}.
The exclusive cross sections of $e^+e^-$ annihilation into open
charm final states could give the detailed coupling strength of these charmonium-like states 
to the open charm final states and could help understanding nature of these charmonium-like states. 
Such exclusive open charm cross sections have been
measured by BaBar~\cite{Aubert:2006mi,Aubert:2009aq,
delAmoSanchez:2010aa} and
Belle~\cite{Pakhlova:2008zza,Abe:2006fj,Pakhlova:2007fq,
Pakhlova:2009jv,Pakhlova:2010ek,Zhukova:2017pen} experiments
through initial states radiation (ISR) and by
CLEO-c~\cite{CroninHennessy:2008yi} experiment with a set of scan
data. The errors of CLEO-c's results are relatively small but
the observed exclusive cross sections are not corrected for the radiative correction. 
As the event selection efficiency in this CLEO-c measurement does not depend on the radiative correction, 
it is possible to apply the radiative correction directly to the observed cross sections to get the
Born cross sections which have clear definition and are more convenient in results comparison and 
theoretical applications.

In this paper, we calculate the ISR correction to the
observed exclusive cross sections of the open charm production in
$e^+e^-$ annihilation measured by
CLEO-c~\cite{CroninHennessy:2008yi}, including the channels
$e^+e^-\to D^0\bar{D}^0$, $D^{*0}\bar{D}^0+c.c.$,
$D^{*0}\bar{D}^{*0}$, $D^+D^-$, $D^{*+}{D}^-+c.c.$,
$D^{*+}{D}^{*-}$, $D^+_s{D}^-_s$, $D^{*+}_s{D}^-_s+c.c.$,
$D^{*+}_s{D}^{*-}_s$, $D^{*}\bar{D}\pi+c.c.$, and
$D^{*}\bar{D}^*\pi$. 
With the calculated ISR correction and the vacuum polarization (VP) correction from
Ref.~\cite{vpfactor}, we obtain the exclusive Born cross sections of the open charm production in $e^+e^-$ annihilation. The
uncertainties of the final Born cross sections from both the
statistical errors of the original measurements and the method to
calculate the ISR correction are estimated properly. By summing up the radiatively corrected exclusive open charm cross sections, we also obtain the Born-level inclusive open charm cross section. 

This paper is organized as follows: the theoretical formulae for
the radiative correction is described first, followed by the
calculation procedure, and at last the final results are
presented.

\section{Radiative correction}

At a center-of-mass energy $\sqrt{s}$, the experimentally observed
cross section ($\sigma^{\mathrm{obs}}$) of $e^+e^-$ annihilation
is the integral of the radiative function $F(x,s)$ and the dressed
cross section
$\sigma^{\mathrm{dre}}(s(1-x))\equiv\sigma^\mathrm{B}(s(1-x))
/|1-\Pi(s(1-x))|^2$~\cite{Kuraev:1985hb},
\begin{align}\label{eq:obsandborn}
\sigma^{\mathrm{obs}}(s)&=\int_0^{x_{\mathrm{m}}}F(x,s)
\sigma^{\mathrm{dre}}(s(1-x))\ \mathrm{d}x\\
&=\int_0^{x_{\mathrm{m}}}F(x,s) \frac{\sigma^{\mathrm{B}}(s(1-x))}
{|1-\Pi(s(1-x))|^2}\ \mathrm{d}x,
\end{align}
where $\sigma^\mathrm{B}(s)$ is the Born cross section and
$1/|1-\Pi(s)|^2$ is the VP factor. The integral variable $x=s'/s$,
where $s'$ is the $e^+e^-$ center of mass energy squared after
emitting ISR photons. The upper limit of the integral
$x_{\mathrm{m}}=1-s_{\mathrm{m}}/s$, where $\sqrt{s_{\mathrm{m}}}$
is the production threshold of a specific channel.

The radiative function $F(x,s)$ is expressed as
\begin{align}
  F(x,s)=&x^{\beta-1}\beta(1+\delta')-\beta(1-\frac{1}{2}x)
 +\frac{1}{2}\beta^2(2-x)\ln\frac{1}{x}\notag\\
 &-\frac{1}{8}\beta^2\left(\frac{(1+3(1-x)^2)}{x}\ln(1-x) -6+x\right)
 \label{eq:F(x,s)}
\end{align}
with
\begin{align}
\beta=&\frac{2\alpha}{\pi}\left(\ln\frac{s}{m_e^2}-1\right),\\
\delta'=&\frac{\alpha}{\pi} \left(\dfrac{\pi^2}{3}-\frac{1}{2}\right)
+\frac{3}{4}\beta+ \beta^2\left(\frac{9}{32}-\frac{\pi^2}{12}\right),
\end{align}
where $\alpha$ is the fine-structure constant and $m_e$ is the
mass of the electron.

To extract the Born cross section from the observed one, we define
\begin{equation}\label{eq:1+delta}
  1+\delta(s)= \sigma^{\mathrm{dre}}(s)/\sigma^{\mathrm{obs}}(s)
\end{equation}
as the ISR correction factor and factorize the Born cross section
as
\begin{equation}\label{eq:bornfactorize}
  \sigma^\mathrm{B}(s)=(1+\delta(s)) \frac{\sigma^{\mathrm{obs}}(s)}
  {1/|1-\Pi(s)|^2}.
\end{equation}

\section{Calculation of ISR correction factors}

\subsection{Methods to connect the data}

To calculate the ISR correction factor, we need full information
of the observed cross section $\sigma^{\mathrm{obs}}$ depending on
the energy square $s$, not only the discrete experimental data.
Five methods to connect the observed cross sections at discrete
energy points (the cross section at the threshold is fixed to zero) are considered and listed below. 
\begin{itemize}
 \item[1.] Linear interpolation, connecting the experimental data
 points with straight lines.
  \item[2.] Cubic spline interpolation with natural boundary
  condition~\cite{csape}. The second derivatives at both ends are zero.
  \item[3.] Cubic smoothing spline~\cite{csaps}.
  \item[4.] B-Spline interpolation of order three~\cite{spapi}.
  \item[5.] B-Spline with uniformly placed knots~\cite{spcrv}.
\end{itemize}
(Only the linear interpolation is considered for $e^+e^-\to D^{*+}_s D^{*-}_s$ and $D^{*}\bar{D}^*\pi$.)

In addition, BaBar and Belle have published the exclusive cross
sections of some channels of charm production in $e^+e^-$
annihilation at the energy region overlapping with that of CLEO-c.
In order to make our results more reliable, for a certain channel
we combine the data of CLEO-c with those of BaBar or Belle and
connect them with linear interpolation as an additional connection
method, as illustrated in Fig.~\ref{fig:belle5}. To be
specific, we consider the data of BaBar~\cite{Aubert:2009aq} for
the channel $e^+e^-\to D^{*0}\bar{D}^0+c.c$,
Belle~\cite{Pakhlova:2008zza} for $D^0\bar{D}^0$, $D^+{D}^-$,
Belle~\cite{Abe:2006fj} for $D^{*+}{D}^{-}+c.c.$,
$D^{*+}{D}^{*-}$, and Belle~\cite{Pakhlova:2010ek} for
$D^+_s{D}^-_s$, $D^{*+}_s{D}^-_s+c.c.$. Since both BaBar and Belle
obtain their results using ISR method, we do not apply the ISR
correction to their data.

 \begin{center}
\includegraphics[width=8cm]{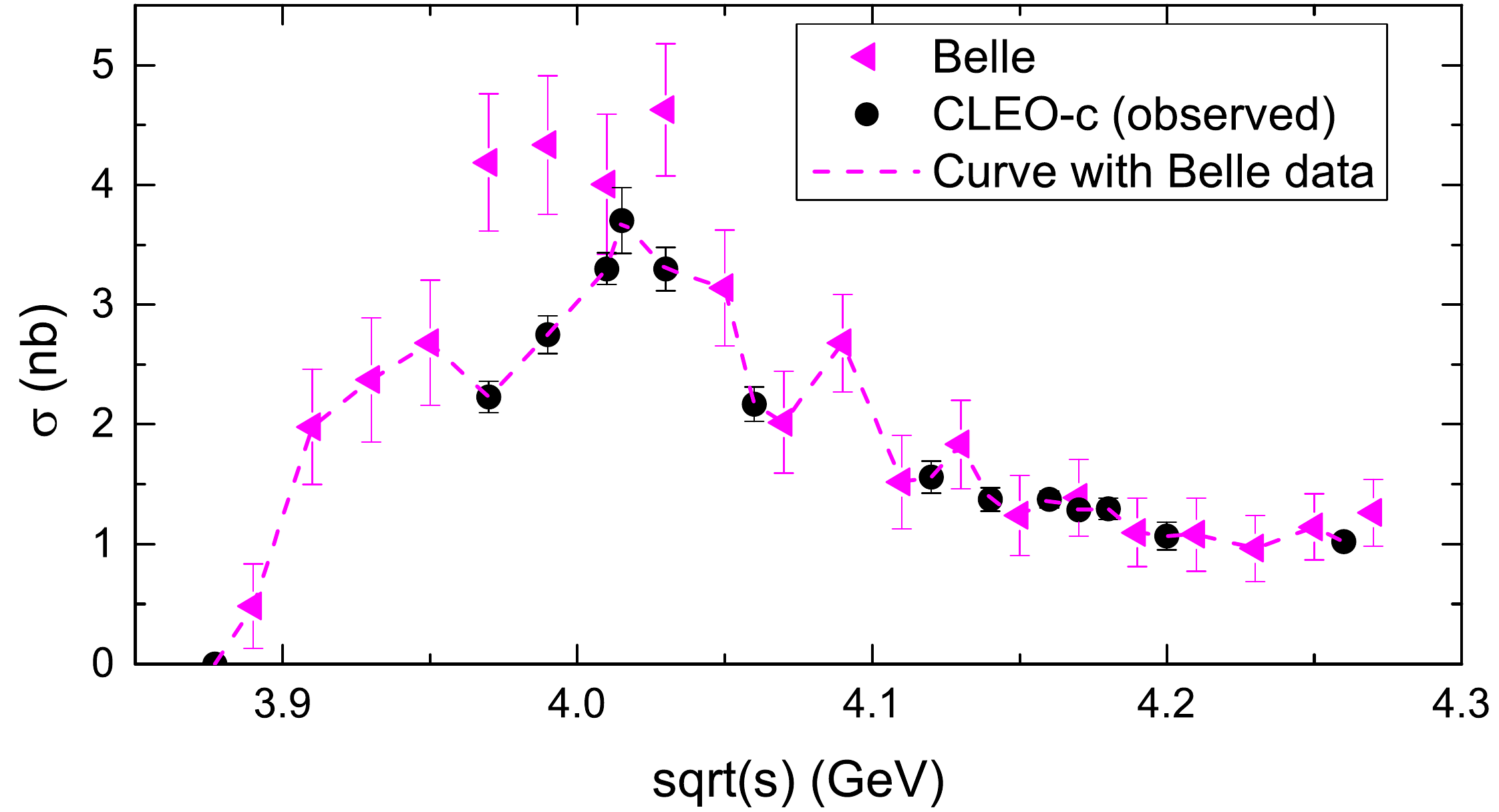}
\figcaption{Illustration of the additional linear connection curve (dashed line) for
$e^+e^-\to D^{*0}\bar{D}^0+c.c.$, where the Belle data Ref.~\cite{Abe:2006fj} 
and CLEO-c data are combined (only statistical errors shown here). 
This curve serves as the initial input of the ISR factor calculation for the additioanl connection method
and then is to be updated within the iterative procedure described later in the text.
}\label{fig:curvewithbelle5}
\end{center}

The comparison of the six curves for the channel $e^+e^-\to
D^{*+}{D}^-+c.c.$ is shown in Fig.~\ref{fig:differentcurvefor5}.

 \begin{center}
\includegraphics[width=8cm]{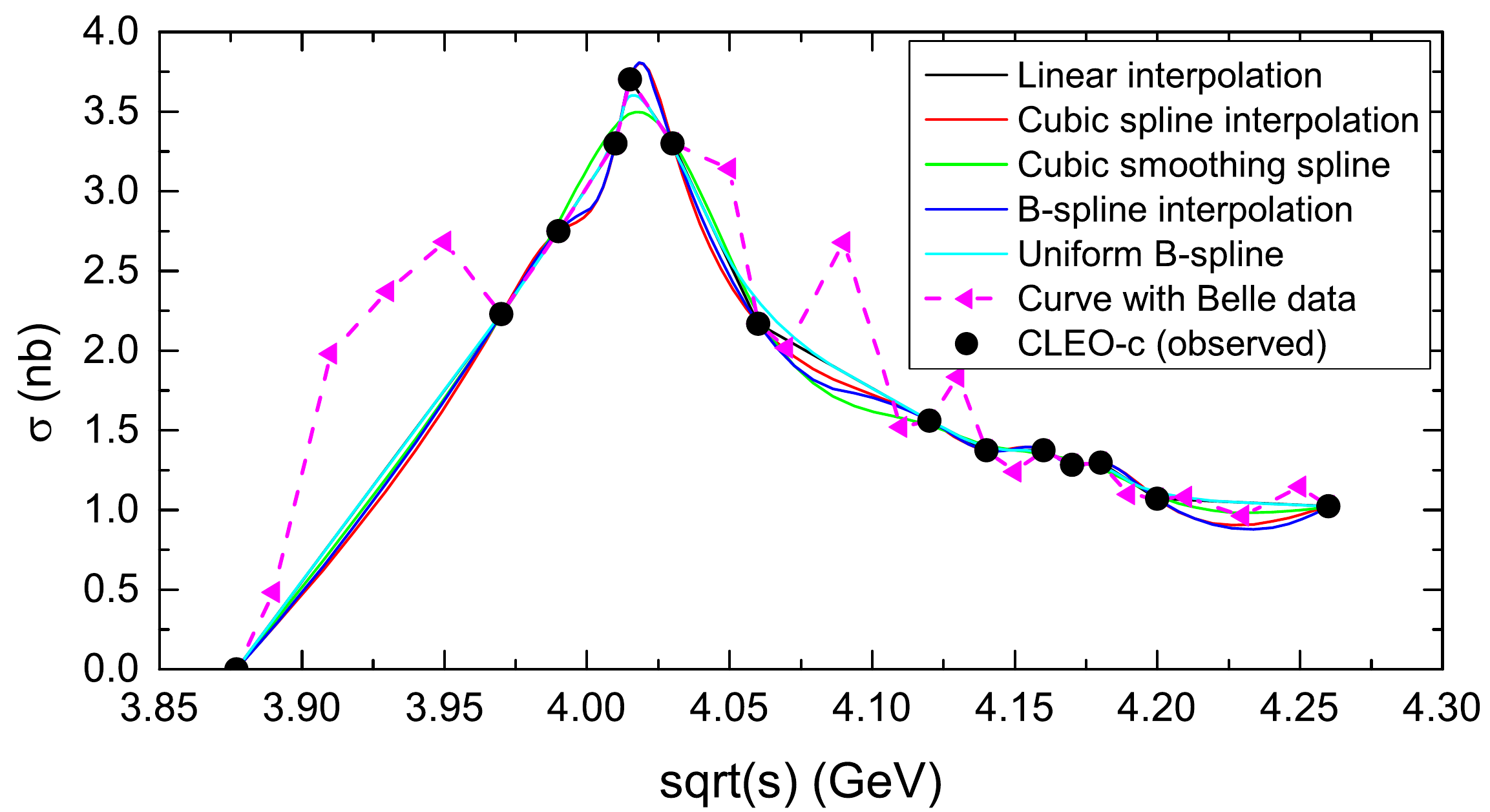}
\figcaption{The connection curves of the cross sections for
$e^+e^-\to D^{*+}{D}^-+c.c.$.  The Belle data are from
Ref.~\cite{Abe:2006fj}.
  }\label{fig:differentcurvefor5}
\end{center}

For the channel $e^+e^-\to D^{0}\bar{D}^{0}$ and $e^+e^-\to
D^{+}{D}^{-}$, the energy points, from $\sqrt{s}=3.970$ to $4.260$
GeV at which CLEO-c~\cite{CroninHennessy:2008yi} experiment is
performed, are far from the production thresholds around 3.74~GeV.
We use the experimental data from Belle~\cite{Pakhlova:2008zza} to
fill the energy gap.

\subsection{Iteration}

The ISR correction factor is obtained in an iterative method via
\begin{align}
  \sigma^{\mathrm{obs}}_{i+1}(s)&=\int_0^{x_{\mathrm{m}}}F(x,s) \sigma^{\mathrm{dre}}_i(s(1-x))\ \mathrm{d}x,\\
  1+\delta_{i+1}(s)&=\sigma^{\mathrm{dre}}_i(s)/ \sigma^{\mathrm{obs}}_{i+1}(s),\\
  \sigma^{\mathrm{dre}}_{i+1}(s)&=(1+\delta_{i+1}(s)) \sigma^{\mathrm{obs}}(s)\label{eq:getsigmadre}
\end{align}
with $\sigma^{\mathrm{dre}}_0(s)= \sigma^{\mathrm{obs}}(s)$. The
iteration is continued until the difference between the two
consecutive results is smaller than the given upper limit, 1\% of
the statistical error of the observation. The results from the
last iteration, denoted by $\sigma^{\mathrm{dre}}_f(s)$ and
$1+\delta_f(s)$, are regarded as the final dressed cross section
and ISR correction factor, respectively.

For example, following the above iteration procedure, we get the
ISR correction factors for $e^+e^-\to D^{*+}{D}^- + c.c.$ with the
linear interpolation and the results converge fast as shown in
Fig.~\ref{fig:iteration5}.

 \begin{center}
\includegraphics[width=8cm]{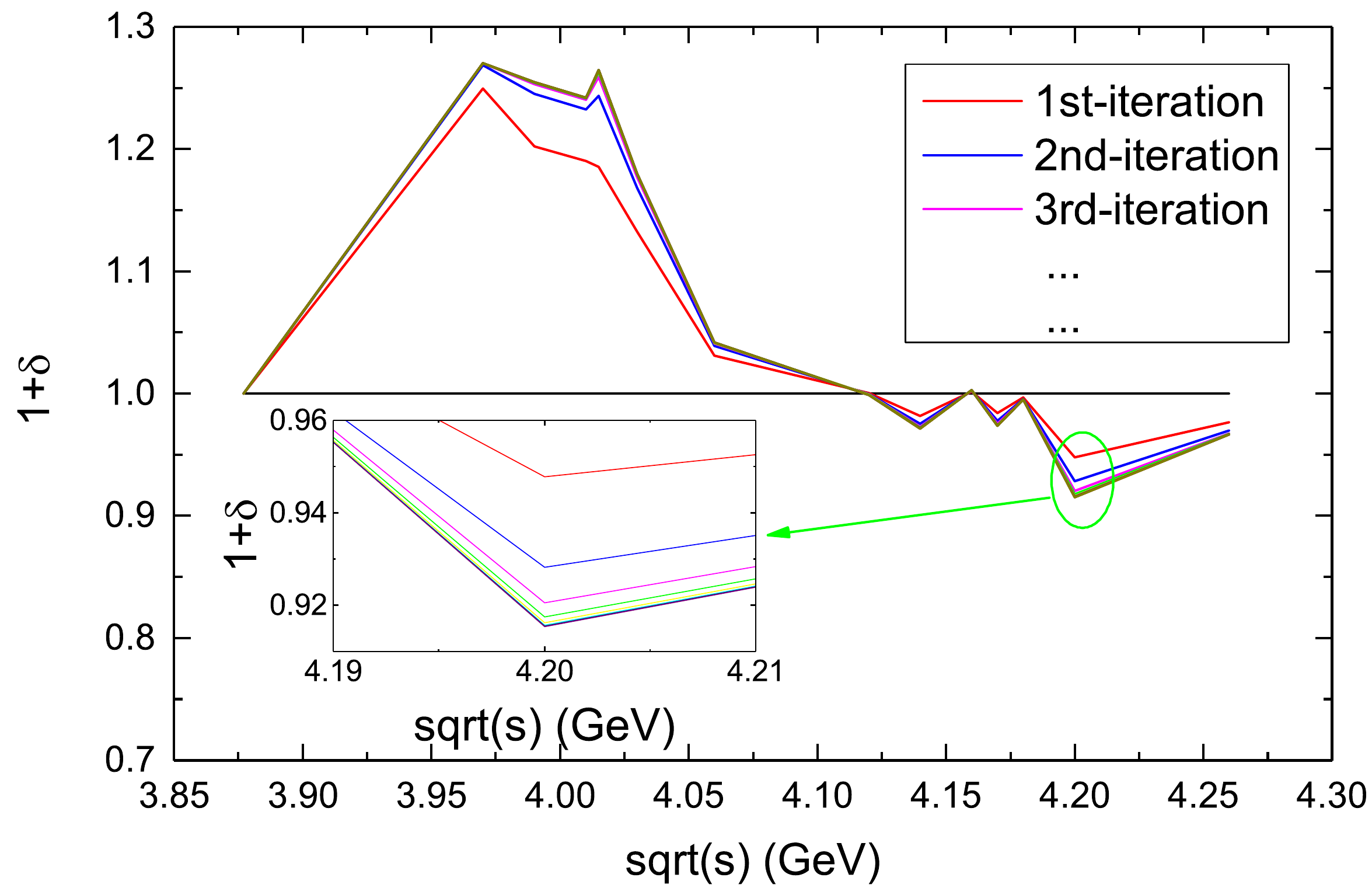}
\figcaption{ The ISR correction factors $1+\delta(s)$ for
$e^+e^-\to D^{*+}{D}^- + c.c.$ with the linear interpolation of the
cross sections after iterations. }\label{fig:iteration5}
\end{center}

\section{Dressed cross sections}
\label{sec:sampling}

We can get the dressed cross sections simultaneously as the ISR
correction factors from the observed cross sections via
Eq.~(\ref{eq:getsigmadre}). Besides the connection methods, the
errors of the observed cross sections also have impacts on the
obtained dressed cross sections. The systematic errors of CLEO-c
experiment are postulated to be the same percentage of the central
values of the observed cross sections (TABLE III in Ref.~\cite{CroninHennessy:2008yi}), which have no modification
to the line shape and in turn the ISR correction factors.
(For $e^+e^-\to D^{0}\bar{D}^{0}$ and
$D^{+}{D}^{-}$, the systematic errors should be treated
in the same way as the statistical errors since Belle data are involved
and their systematic errors are independent of those of CLEO-c
data.) While considering the statistical errors of the observed
cross sections which are independent between energy points, the
line shape and the obtained dressed cross sections could be
different. To investigate the effects of the statistical errors of
the observation on the dressed cross sections, we perform
samplings of the $\sigma^{\mathrm{dre}}_0(s)$ according to a
Gaussian distribution at each energy point, where the mean value
and the standard deviation of the Gaussian distribution are the
central value and the statistical error of the observed cross
section, respectively. (The negative cross sections from the
samplings are discarded.) Following the iteration procedure with a
certain connection method, we obtain the dressed cross sections
for each sampling.

After 30,000 samplings we get a distribution of the dressed cross
section at each energy point. We find that most results satisfy
Gaussian distribution so the fitted mean and standard deviation
are taken as the central value and the statistical error of the
dressed cross section. However, at some energy point the observed cross section is so small that it is possible to get some sampled values which yield very small (going to zero during iterations) ISR factors or equivalently very small dressed cross sections. It also indicates that at these energy points all the observed events could be from ISR. For these results, if there is a small amount of 
$\sigma^{\mathrm{dre}}_f(s)$ accumulating at zero, 
we ignore it and still try to fit the sampling results with
Gaussian distributions (Fig.~\ref{fig:nongauss}(a)), 
and if the accumulation at zero dominates, we try to estimate the
upper limit for the dressed cross section
(Fig.~\ref{fig:nongauss}(b)).

 \begin{center}
\includegraphics[width=9cm]{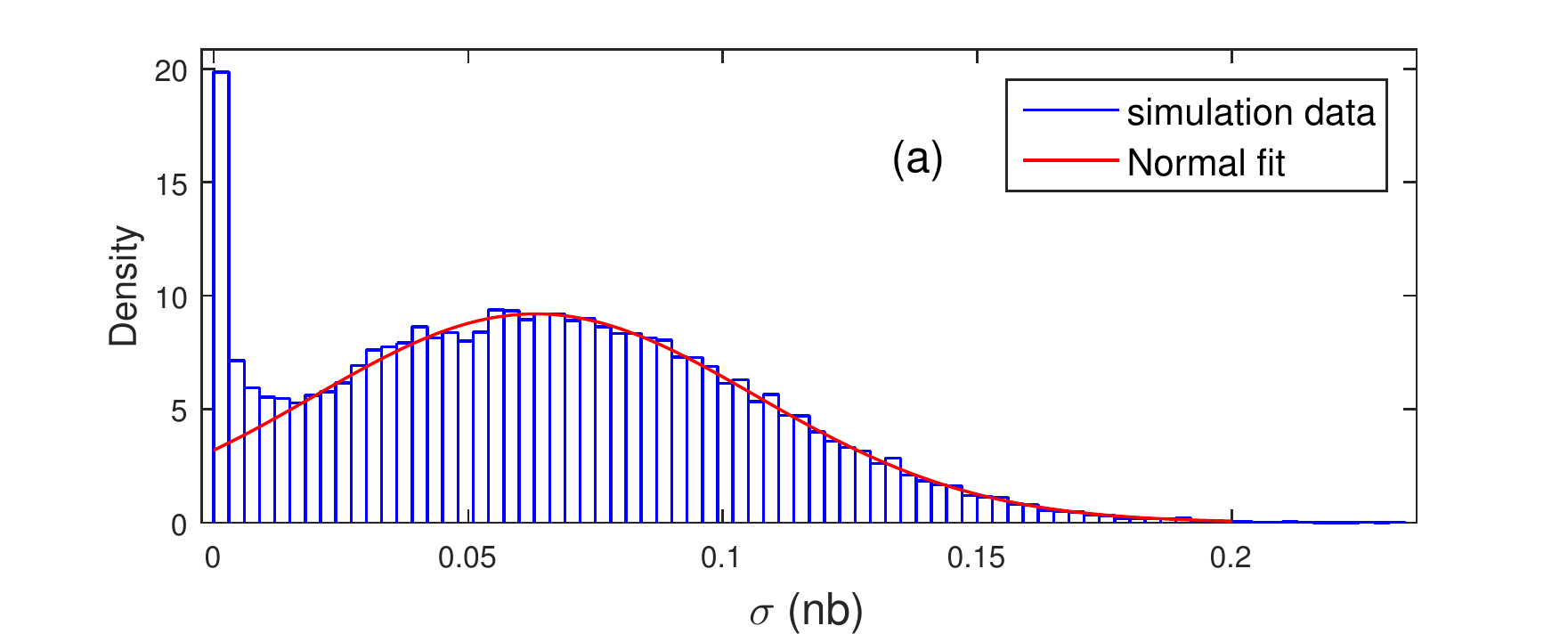}\\
\includegraphics[width=9cm]{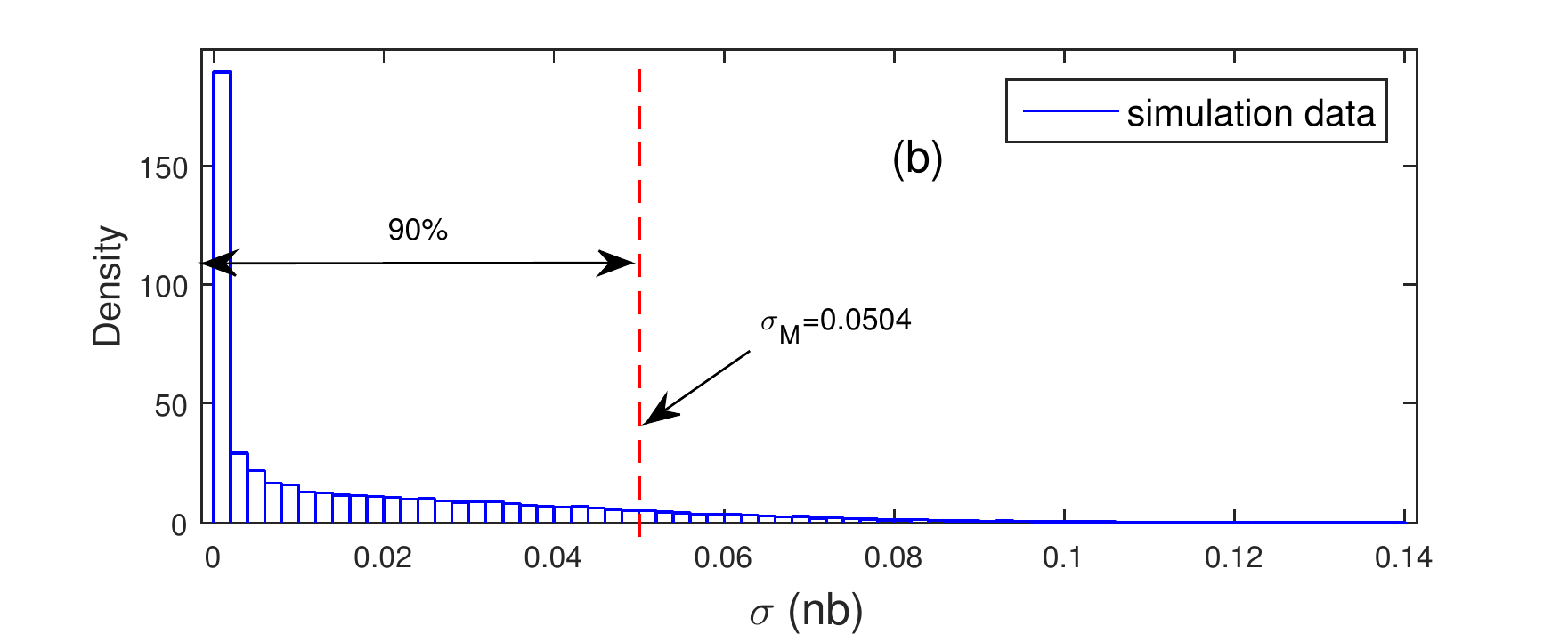}
\figcaption{ Examples of simulation results not satisfying
Gaussian distribution. It's acceptable to describe (a) with a truncated Gaussian distribution while for (b) we only estimate an upper limit. }\label{fig:nongauss}
\end{center}
 \begin{center}
\includegraphics[width=8cm]{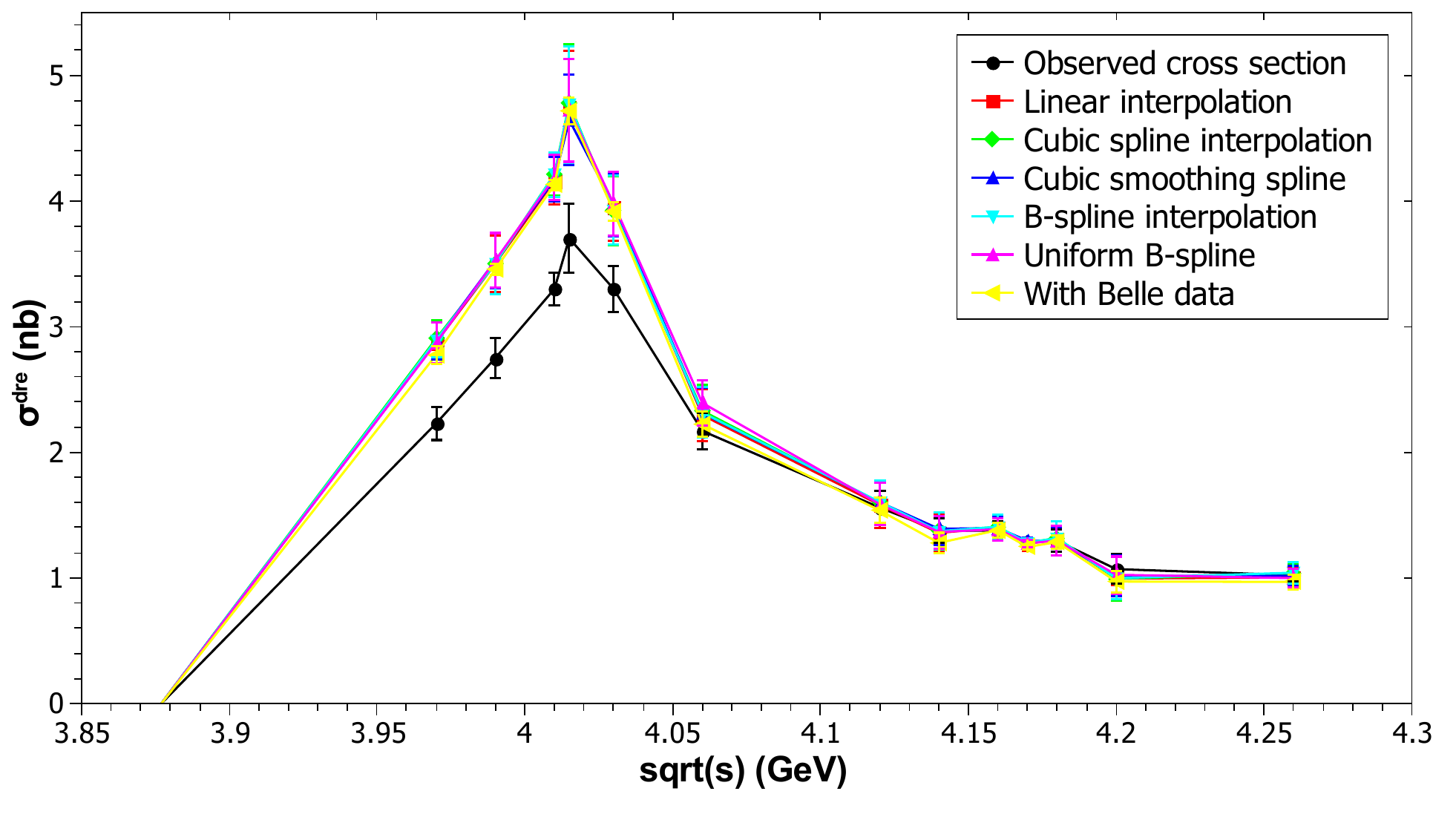}
\figcaption{ The dressed cross sections from different connection
methods for $e^+e^-\to D^{*+}{D}^-+c.c.$ }\label{fig:totalsigma5}
\end{center}

 Different results are obtained after applying the calculation
procedure with different connection methods, denoted by $\sigma^{\mathrm{dre}}_i=\alpha_i\pm\epsilon_{\mathrm{sta},i}$ where $\alpha_i$ and $\epsilon_{\mathrm{sta},i}$ are the central value and the statistical error from the $i$-th connection method. 
As an example, the results with different connection methods for the channel
$e^+e^-\to D^{*+}{D}^-+c.c.$ are illustrated in
Fig.~\ref{fig:totalsigma5}. 
The final dressed cross section is
expressed in the form of
 $\sigma^{\mathrm{dre}}=\alpha\pm
  \epsilon_{\mathrm{sta}}\
 ^{+\epsilon_{\mathrm{sys}}^{\mathrm{ISR}+}}
 _{-\epsilon_{\mathrm{sys}}^{\mathrm{ISR}-}}$
(the systematic error of CLEO-c measurements is not included
here) where the central value $\alpha$ and the statistical error
$\epsilon_{\mathrm{sta}}$ are from the linear interpolation
method, and the systematic error $\epsilon_{\mathrm{sys}}^{\mathrm{ISR}\pm}$
is derived from the total error $\epsilon_{\mathrm{tot}}^{\mathrm{ISR}\pm}$ and the statistical error $\epsilon_{\mathrm{sta}}$ by 
$\epsilon_{\mathrm{sys}}^{\mathrm{ISR}\pm} \equiv
\sqrt{(\epsilon_{\mathrm{tot}}^{\mathrm{ISR}\pm})^2-
(\epsilon_{\mathrm{sta}}^{\pm})^2}$.
At each energy point the upper/lower value of the total
error from the ISR correction is estimated by 
\begin{align}
\epsilon_{\mathrm{tot}}^{\mathrm{ISR}+}&=\mathrm{max}(\{\alpha_i+\epsilon_{\mathrm{sta},i}-\alpha\}),\\
\epsilon_{\mathrm{tot}}^{\mathrm{ISR}-}&=\mathrm{max}(\{\alpha-\alpha_i+\epsilon_{\mathrm{sta},i}\})
\end{align}
where max(\{$x_i$\}) yields the maximum one among $x_i$.
Note that $\epsilon_{\mathrm{sys}}^{\mathrm{ISR}\pm}$ is just an estimation of the uncertainty from connection methods. For the simulation results described by the 90\% upper confidence limit, we just take the biggest upper limit among these different connection methods as the result considering the systematic uncertainties from the correction methods.



\section{VP factors and Born cross sections}

To obtain the Born cross sections, we divide the dressed cross
sections by the VP factors $1/|1-\Pi(s)|^2$, which are calculated
in Ref.~\cite{vpfactor} and listed in Appendix~\ref{a1}. The
obtained exclusive Born cross sections are shown in
Fig.~\ref{fig:finalsigma} and listed in Appendix~\ref{a2}.

 \begin{center}
\includegraphics[width=8cm]{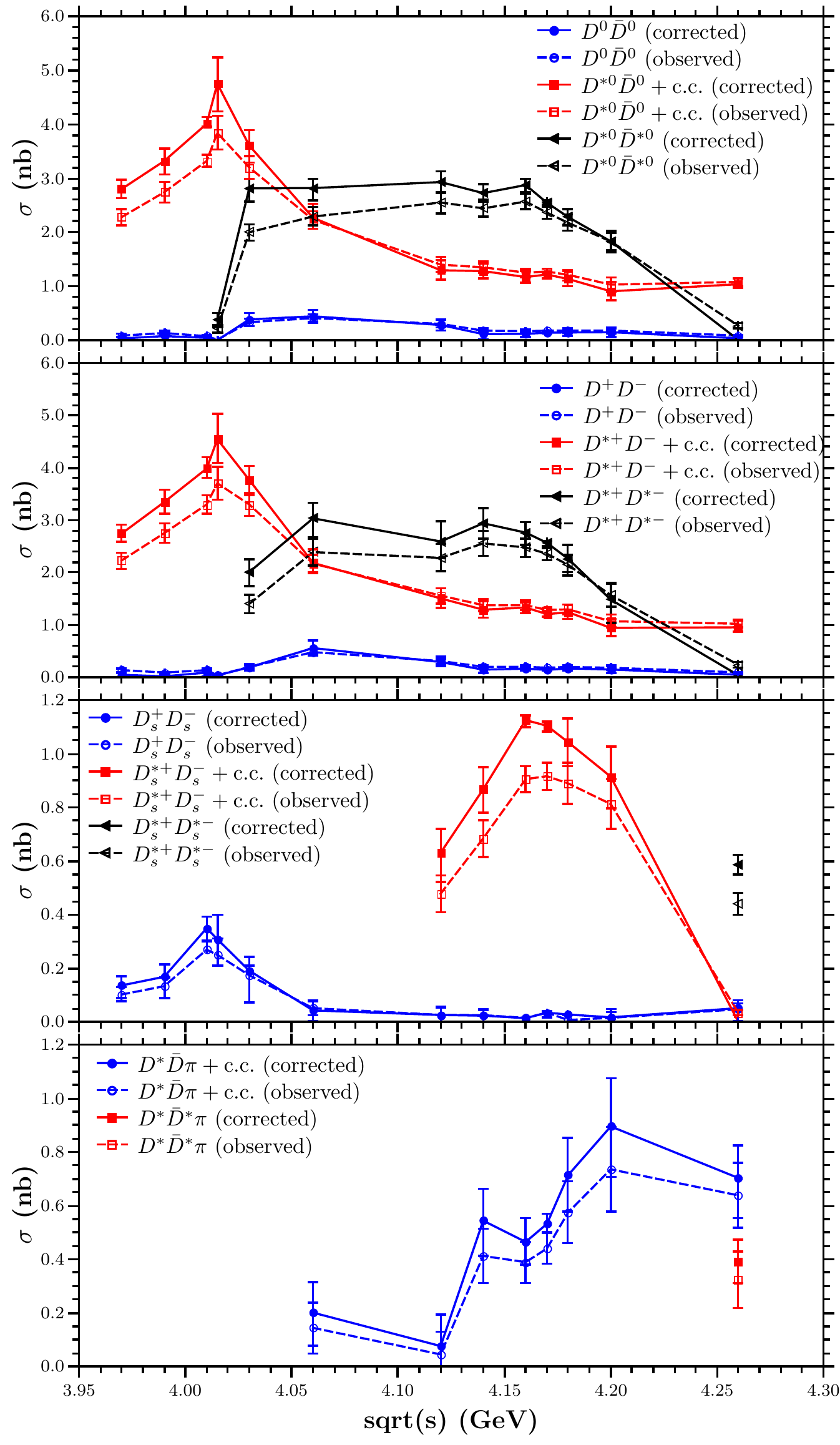}
\figcaption{ Born (corrected) and observed cross sections for the exclusive open charm
meson productions in $e^+e^-$ annihilation with combined
statistical and systematic (only $\epsilon_{\mathrm{sys}}^{\mathrm{ISR}\pm}$ for the corrected cross sections) errors. 
}\label{fig:finalsigma}
\end{center}
 \begin{center}
\includegraphics[width=8cm]{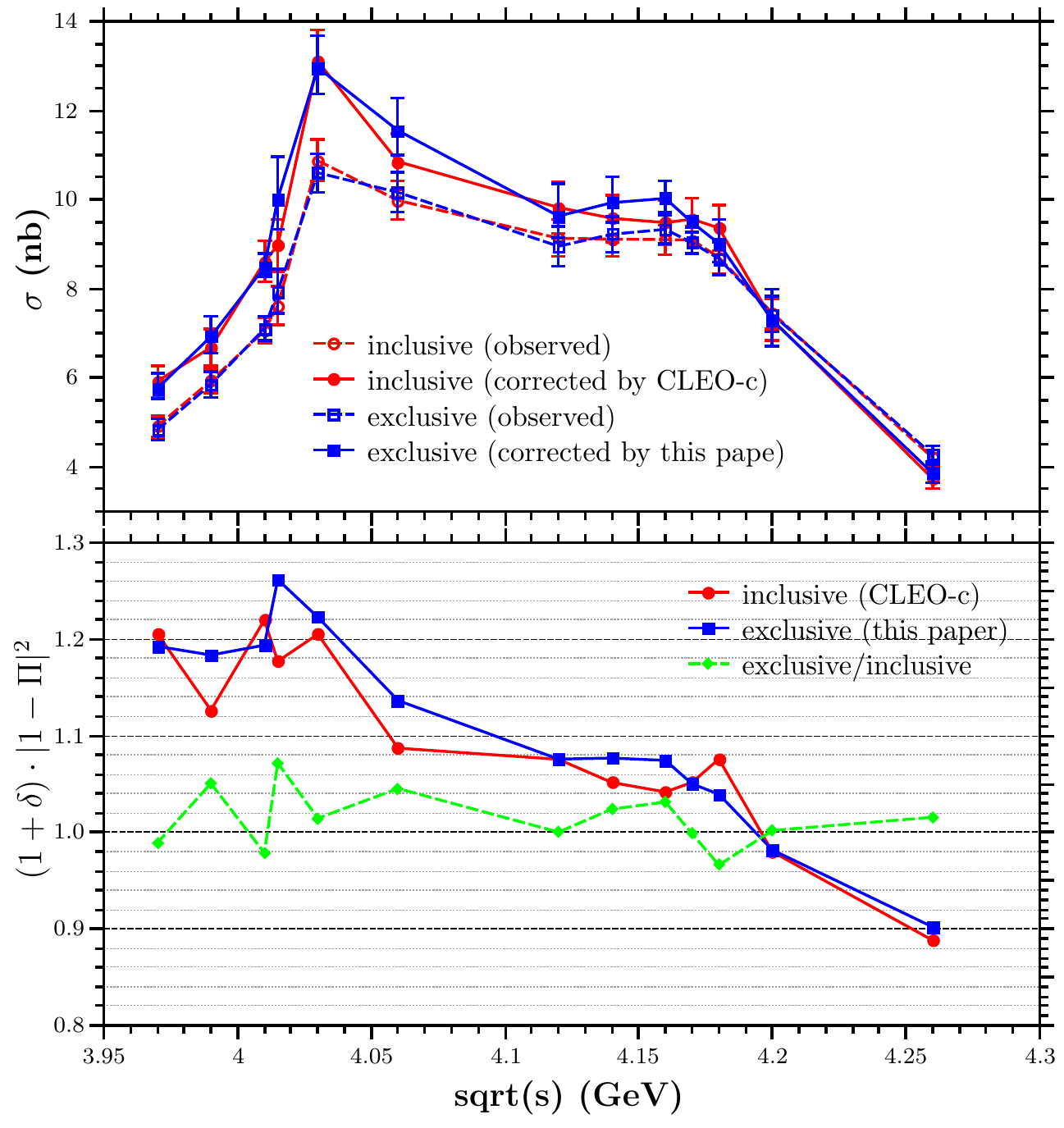}
\figcaption{  The comparison between the radiative correction performed by CLEO-c and this study. The ``exclusive" curves represent the summation of exclusive cross sections. In Ref.~\cite{CroninHennessy:2008yi}, the total charm cross sections determined by summing up the exclusive measurements (blue dashed line) and by inclusive charm counting method (red dashed line) are not exactly the same. So it is clearer to compare the nominal overall radiative correction factors, showing in the second figure here.
}\label{fig:totcompare_cleoc}
\end{center}

Summing up all the exclusive cross sections in Fig.~\ref{fig:finalsigma} we can obtained the inclusive Born cross section. 
At most of the energy points the exclusive results have symmetric statistical errors so the summing up is straight forward. 
For some energy points, part of the exclusive results have asymmetric statistical errors. In such cases, we try to obtain the summed results also through the toy MC method. We sum the exclusive cross sections up after a certain sampling for each channel and get distributions for the summed cross section after the 30,000 samplings (see section~\ref{sec:sampling}). These distributions are quite symmetric and can be fitted by the Gaussian distribution to yield the central value $\sigma^{\mathrm{B}}_{\mathrm{inc}}$ and the statistical error $\epsilon_{\mathrm{sta,inc}}^{\mathrm{ISR}}$.
The systematic errors from ISR for the summation are all estimated by 
\begin{equation}\label{eq:sumoferr}
\epsilon_{\mathrm{sys, inc}}^{\mathrm{ISR}\pm}=\sqrt{\sum_i(\epsilon_{\mathrm{sys},i}^{\mathrm{ISR}\pm})^2},
\end{equation} 
assuming the systematic errors of different exclusive channels are independent, where $i$ is the index of exclusive channels. As mentioned in Section 4, the systematic errors of the original measurement are assumed to be the same percentage of the central values within a certain channel. For the Born cross section, the same percentage of the central values are taken directly as part of the systematic errors for this channel, denoted by $\epsilon_{\mathrm{sys},i}^{\mathrm{CLEOc}}$. Assuming that this part of systematic errors for different channels are independent, they are summed up in the same way as Eq.(\ref{eq:sumoferr}) to yield this part of the systematic error of the inclusive Born cross section, denoted by $\epsilon_{\mathrm{sys,inc}}^{\mathrm{CLEOc}}$. Put what we obtained above together and we get the final results for the inclusive Born cross section for $e^+e^-$ annihilation into open charm final states, which are listed in Appendix~\ref{a1}.

The comparison between the inclusive Born cross section obtained by this study and by CLEO-c is illustrated in Fig.~\ref{fig:totcompare_cleoc}. The top figure shows the cross sections and the bottom one shows the differences between the nominal overall radiative correction factors $(1+\delta)\cdot|1-\Pi|^2=\sigma^{\mathrm{B}}/\sigma^{\mathrm{obs}}$. No significant difference is observed.
In principle, it is more reasonable to obtain the inclusive cross section by summing up the exclusive ones because by treating each channel separately the individual open-charm production thresholds and different exclusive cross section line shapes can be implemented in the ISR correction.

\section{Summary and discussion}

In summary, we have obtained the Born cross sections for charm
production in $e^+e^-$ annihilation after correcting the observed
cross sections measured by CLEO-c experiment for initial states
radiation and vacuum polarization. A Monte Carlo sampling method
is used to estimate the uncertainty of the corrected cross
sections caused by the original statistical errors. The
uncertainty from the connection of the observations at discrete
energy points is estimated with different connection methods when
we determine the errors of the Born cross sections and regarded as
a part of the systematic error. It turns out that such systematic
errors are of the same order as the statistical ones. In addition,
we make our results more reliable by taking the corresponding data
obtained by Belle and BaBar collaboration into consideration.

The Born cross sections for the exclusive open charm meson productions in $e^+e^-$ annihilation, which are from the observation at CLEO-c corrected for ISR and VP effects, can provide more conventional information to determine the relative decay strength into different charm mesons for the charmonium(-like) states in the open charm region and can be easily compared or combined with the results from other experiments. Besides, the summation of the radiatively corrected exclusive cross sections, which is more reliable than the one obtain by correcting the observed inclusive cross section directly, can be used in the calculation of the anomalous magnetic moment of muon $g-2$ and the vacuum polarization factor $1/|1-\Pi(s)|^2$.


\end{multicols}

\vspace{-1mm}
\centerline{\rule{80mm}{0.1pt}}
\vspace{2mm}

\begin{multicols}{2}

\end{multicols}

\vspace{10mm}
\newpage

\appendix
\section{Inclusive Born cross sections and vacuum polarization factors}
\label{a1}

\begin{small}
\begin{center}
\tabcaption{Inclusive Born cross sections and the VP factors (from
Ref.~\cite{vpfactor}) at the 13 energy points.
$\sigma^{\mathrm{B}}_{\mathrm{inc}}$ is the inclusive Born cross section. The following four columns are different types of errors of the inclusive Born cross section defined in the text.
 The last column is the VP factor.}\label{tab:vpfactor}
\begin{tabular}{ccccccc}
\toprule
$\sqrt{s}$(GeV) & $\sigma^{\mathrm{B}}_{\mathrm{inc}}$ (nb) & $\epsilon_{\mathrm{sta,inc}}^{\mathrm{ISR}} $ (nb) & $\epsilon_{\mathrm{sys,inc}}^{\mathrm{ISR}+} $ (nb) & $\epsilon_{\mathrm{sys,inc}}^{\mathrm{ISR}-} $ (nb) & $\epsilon_{\mathrm{sys,inc}}^{\mathrm{CLEOc}} $ (nb) & $1/|1-\Pi(s)|^2$ \\
\hline
3.970  & 5.73  & 0.24  & 0.25  & 0.25  & 0.14  & 1.0480  $\pm$ 0.0009  \\
    3.990  & 6.92  & 0.32  & 0.35  & 0.37  & 0.17  & 1.0472  $\pm$ 0.0008  \\
    4.010  & 8.45  & 0.20  & 0.26  & 0.22  & 0.20  & 1.0445  $\pm$ 0.0008  \\
    4.015  & 9.99  & 0.64  & 0.71  & 0.70  & 0.23  & 1.0448  $\pm$ 0.0008  \\
    4.030  & 12.96  & 0.50  & 0.52  & 0.59  & 0.25  & 1.0492  $\pm$ 0.0008  \\
    4.060  & 11.54  & 0.48  & 0.56  & 0.56  & 0.23  & 1.0517  $\pm$ 0.0009  \\
    4.120  & 9.64  & 0.47  & 0.55  & 0.49  & 0.20  & 1.0508  $\pm$ 0.0007  \\
    4.140  & 9.93  & 0.41  & 0.44  & 0.44  & 0.22  & 1.0524  $\pm$ 0.0007  \\
    4.160  & 10.01  & 0.29  & 0.29  & 0.33  & 0.22  & 1.0527  $\pm$ 0.0007  \\
    4.170  & 9.48  & 0.09  & 0.10  & 0.13  & 0.20  & 1.0546  $\pm$ 0.0007  \\
    4.180  & 8.99  & 0.40  & 0.40  & 0.41  & 0.20  & 1.0553  $\pm$ 0.0007  \\
    4.200  & 7.29  & 0.49  & 0.50  & 0.60  & 0.17  & 1.0564  $\pm$ 0.0007  \\
    4.260  & 3.81  & 0.19  & 0.21  & 0.21  & 0.15  & 1.0521  $\pm$ 0.0006  \\

\bottomrule
\end{tabular}
\end{center}

\end{small}

\section{Exclusive Born cross sections}
\label{a2}

\begin{center}
\begin{small}
\tabcaption{ \label{tab:sigmaD} The Born cross sections for
final states consisting of two neutral nonstrange charm mesons.
The first error of each cross section is statistical and the
second is systematic ($\epsilon_{\mathrm{sys}}^{\mathrm{ISR}}$). The upper limits are at 90\% confidence
level.}

\begin{tabular}{cccc}
\toprule
$\sqrt{s}$ (GeV) & $\sigma^B(D^{0}\bar{D}^0)$(nb) & $\sigma^B(D^{*0}\bar{D}^0+c.c.)$(nb)& $\sigma^B(D^{*0}\bar{D}^{*0})$(nb)\\
\hline
3.970 & $<$ 0.033 & 2.800 $\pm\ 0.165^{+ 0.020 }_{- 0.001}$ & $\cdots$ \\
    3.990 & 0.076 $\pm\ 0.056^{+ 0.001 }_{- 0.072}$  & 3.319 $\pm\ 0.218^{+ 0.091 }_{- 0.130}$  & $\cdots$ \\
    4.010 &$<$ 0.044 & 4.007 $\pm\ 0.032^{+ 0.118 }_{- 0.035}$  & $\cdots$ \\
    4.015 &$<$ 0.005 & 4.752 $\pm\ 0.442^{+ 0.212 }_{- 0.253}$  & 0.377 $\pm\ 0.132^{+ 0.001 }_{- 0.028}$  \\
    4.030 & 0.385 $\pm\ 0.102^{+ 0.040 }_{- 0.075}$  & 3.605 $\pm\ 0.262^{+ 0.114 }_{- 0.156}$  & 2.813 $\pm\ 0.178^{+ 0.001 }_{- 0.165}$  \\
    4.060 & 0.443 $\pm\ 0.100^{+ 0.073 }_{- 0.072}$  & 2.268 $\pm\ 0.199^{+ 0.161 }_{- 0.001}$  & 2.817 $\pm\ 0.179^{+ 0.001 }_{- 0.149}$  \\
    4.120 & 0.283 $\pm\ 0.091^{+ 0.046 }_{- 0.061}$  & 1.296 $\pm\ 0.173^{+ 0.075 }_{- 0.001}$  & 2.929 $\pm\ 0.198^{+ 0.001 }_{- 0.001}$  \\
    4.140 & 0.114 $\pm\ 0.057^{+ 0.024 }_{- 0.072}$  & 1.280 $\pm\ 0.142^{+ 0.057 }_{- 0.001}$  & 2.725 $\pm\ 0.164^{+ 0.033 }_{- 0.024}$  \\
    4.160 & 0.120 $\pm\ 0.041^{+ 0.001 }_{- 0.052}$  & 1.171 $\pm\ 0.098^{+ 0.040 }_{- 0.001}$  & 2.874 $\pm\ 0.118^{+ 0.001 }_{- 0.056}$  \\
    4.170 & 0.139 $\pm\ 0.016^{+ 0.001 }_{- 0.021}$  & 1.217 $\pm\ 0.033^{+ 0.026 }_{- 0.076}$  & 2.546 $\pm\ 0.035^{+ 0.018 }_{- 0.030}$  \\
    4.180 & 0.144 $\pm\ 0.058^{+ 0.001 }_{- 0.032}$  & 1.134 $\pm\ 0.135^{+ 0.040 }_{- 0.001}$  & 2.280 $\pm\ 0.153^{+ 0.001 }_{- 0.047}$  \\
    4.200 & 0.149 $\pm\ 0.075^{+ 0.012 }_{- 0.033}$  & 0.906 $\pm\ 0.170^{+ 0.060 }_{- 0.001}$  & 1.831 $\pm\ 0.193^{+ 0.001 }_{- 0.062}$  \\
    4.260 & 0.036 $\pm\ 0.024^{+ 0.006 }_{- 0.024}$  & 1.038 $\pm\ 0.077^{+ 0.087 }_{- 0.029}$  & $<$ 0.006  \\
\bottomrule
\end{tabular}%
\end{small}
\end{center}

\newpage
\begin{center}
\begin{small}
\tabcaption{ \label{tab:sigmaD+} The Born cross sections for
final states consisting of two charged nonstrange mesons. The
first error of each cross section is statistical and the second is
systematic ($\epsilon_{\mathrm{sys}}^{\mathrm{ISR}}$). The upper limits are at 90\% confidence level.}

\begin{tabular}{cccc}
\toprule
$\sqrt{s}$ (GeV) & $\sigma^B(D^{+}{D}^-)$(nb) & $\sigma^B(D^{*+}{D}^-+c.c.)$(nb)& $\sigma^B(D^{*+}{D}^{*-})$(nb)\\
\hline
    3.970 & 0.048 $\pm\ 0.040 ^{+ 0.001 }_{- 0.025}$ & 2.743 $\pm\ 0.162 ^{+ 0.049 }_{- 0.020}$ & $\cdots$ \\
    3.990 & 0.017 $\pm\ 0.030 ^{+ 0.022 }_{- 0.001}$ & 3.343 $\pm\ 0.219 ^{+ 0.088 }_{- 0.067}$ & $\cdots$ \\
    4.010 & 0.090 $\pm\ 0.033 ^{+ 0.001 }_{- 0.069}$ & 3.985 $\pm\ 0.186 ^{+ 0.107 }_{- 0.001}$ & $\cdots$ \\
    4.015 & $<$ 0.035 & 4.545 $\pm\ 0.427 ^{+ 0.210 }_{- 0.133}$ & $\cdots$ \\
    4.030 & 0.190 $\pm\ 0.052 ^{+ 0.031 }_{- 0.028}$ & 3.769 $\pm\ 0.260 ^{+ 0.037 }_{- 0.134}$ & 2.007 $\pm\ 0.245 ^{+ 0.069 }_{- 0.098}$ \\
    4.060 & 0.558 $\pm\ 0.079 ^{+ 0.111 }_{- 0.057}$ & 2.183 $\pm\ 0.194 ^{+ 0.183 }_{- 0.001}$ & 3.034 $\pm\ 0.298 ^{+ 0.018 }_{- 0.210}$ \\
    4.120 & 0.291 $\pm\ 0.066 ^{+ 0.088 }_{- 0.038}$ & 1.506 $\pm\ 0.174 ^{+ 0.055 }_{- 0.001}$ & 2.593 $\pm\ 0.296 ^{+ 0.238 }_{- 0.107}$ \\
    4.140 & 0.146 $\pm\ 0.043 ^{+ 0.022 }_{- 0.034}$ & 1.291 $\pm\ 0.138 ^{+ 0.072 }_{- 0.060}$ & 2.940 $\pm\ 0.274 ^{+ 0.104 }_{- 0.103}$ \\
    4.160 & 0.164 $\pm\ 0.031 ^{+ 0.019 }_{- 0.015}$ & 1.331 $\pm\ 0.098 ^{+ 0.001 }_{- 0.025}$ & 2.762 $\pm\ 0.190 ^{+ 0.001 }_{- 0.140}$ \\
    4.170 & 0.141 $\pm\ 0.014 ^{+ 0.013 }_{- 0.010}$ & 1.205 $\pm\ 0.032 ^{+ 0.041 }_{- 0.034}$ & 2.566 $\pm\ 0.043 ^{+ 0.001 }_{- 0.032}$ \\
    4.180 & 0.169 $\pm\ 0.041 ^{+ 0.018 }_{- 0.015}$ & 1.241 $\pm\ 0.128 ^{+ 0.042 }_{- 0.001}$ & 2.260 $\pm\ 0.256 ^{+ 0.001 }_{- 0.054}$ \\
    4.200 & 0.148 $\pm\ 0.051 ^{+ 0.021 }_{- 0.022}$ & 0.942 $\pm\ 0.161 ^{+ 0.057 }_{- 0.028}$ & 1.482 $\pm\ 0.300 ^{+ 0.001 }_{- 0.319}$ \\
    4.260 & 0.048 $\pm\ 0.018 ^{+ 0.006 }_{- 0.046}$ & 0.954 $\pm\ 0.071 ^{+ 0.093 }_{- 0.049}$ & $<$ 0.038 \\
\bottomrule
\end{tabular}%
\end{small}
\end{center}

\begin{center}
\begin{small}
\tabcaption{ \label{tab:sigmaDs} The Born cross sections for final states consisting of two strange charm mesons. The first error of each cross section is statistical and the second is systematic ($\epsilon_{\mathrm{sys}}^{\mathrm{ISR}}$). The upper limits are at 90\% confidence level. The systematic error ($\epsilon_{\mathrm{sys}}^{\mathrm{ISR}}$) for $e^+e^-\to D^{*+}_s\bar{D}^{*-}_s$ is not estimated so only the statistical one is presented.}

\begin{tabular}{cccc}
\toprule
$\sqrt{s}$(GeV) & $\sigma^B(D^{+}_s{D}^-_s)$(nb) & $\sigma^B(D^{*+}_s{D}^-_s)$(nb)& $\sigma^B(D^{*+}_s{D}^{*-}_s)$(nb)\\
\hline
    3.970 & 0.136 $\pm\ 0.035 ^{+ 0.001 }_{- 0.031}$ & $\cdots$ & $\cdots$ \\
    3.990 & 0.168 $\pm\ 0.043 ^{+ 0.015 }_{- 0.067}$ & $\cdots$ & $\cdots$ \\
    4.010 & 0.349 $\pm\ 0.042 ^{+ 0.001 }_{- 0.025}$ & $\cdots$ & $\cdots$ \\
    4.015 & 0.308 $\pm\ 0.092 ^{+ 0.012 }_{- 0.034}$ & $\cdots$ & $\cdots$ \\
    4.030 & 0.191 $\pm\ 0.052 ^{+ 0.007 }_{- 0.107}$ & $\cdots$ & $\cdots$ \\
    4.060 & 0.043 $\pm\ 0.032 ^{+ 0.009 }_{- 0.020}$ & $\cdots$ & $\cdots$ \\
    4.120 & 0.026 $\pm\ 0.029 ^{+ 0.010 }_{- 0.001}$ & 0.630 $\pm\ 0.085 ^{+ 0.020 }_{- 0.066}$ & $\cdots$ \\
    4.140 & 0.023 $\pm\ 0.025 ^{+ 0.006 }_{- 0.001}$ & 0.867 $\pm\ 0.082 ^{+ 0.015 }_{- 0.026}$ & $\cdots$ \\
    4.160 & $<$ 0.014  & 1.125 $\pm\ 0.018 ^{+ 0.001 }_{- 0.017}$ & $\cdots$ \\
    4.170 & 0.034 $\pm\ 0.005 ^{+ 0.001 }_{- 0.016}$ & 1.104 $\pm\ 0.017^{+ 0.002 }_{- 0.011}$ & $\cdots$ \\
    4.180 & $<$ 0.027 & 1.042 $\pm\ 0.087 ^{+ 0.001 }_{- 0.015}$ & $\cdots$ \\
    4.200 & 0.017 $\pm\ 0.026 ^{+ 0.015 }_{- 0.001}$ & 0.914 $\pm\  0.114^{+ 0.001 }_{- 0.034}$ & $\cdots$\\
    4.260 & 0.051 $\pm\ 0.027 ^{+ 0.008 }_{- 0.041}$ & $<$ 0.001 & 0.585 $\pm$ 0.036 \\

\bottomrule
\end{tabular}%
\end{small}
\end{center}

\begin{center}
\begin{small}
\tabcaption{ \label{tab:sigmamulti} The Born cross sections for
final states consisting of two charm mesons and an extra pion. The
first error of each cross section is statistical and the second is
systematic ($\epsilon_{\mathrm{sys}}^{\mathrm{ISR}}$). The systematic error ($\epsilon_{\mathrm{sys}}^{\mathrm{ISR}}$) for $e^+e^-\to D^{*}\bar{D}^*\pi$ is not estimated so only the statistical one is presented.}

\begin{tabular}{ccc}
\toprule
$\sqrt{s}$ (GeV) & $\sigma^B(D^{*}\bar{D}\pi+c.c.)$(nb) & $\sigma^B(D^{*}\bar{D}^*\pi)$(nb)\\
\hline
    4.060 & 0.201 $\pm\ 0.114^{+ 0.001 }_{- 0.048}$ & $\cdots$ \\
    4.120 & 0.076 $\pm\ 0.090^{+ 0.076 }_{- 0.050}$ & $\cdots$ \\
    4.140 & 0.544 $\pm\ 0.121^{+ 0.001 }_{- 0.052}$ & $\cdots$ \\
    4.160 & 0.464 $\pm\ 0.084^{+ 0.034 }_{- 0.001}$ & $\cdots$ \\
    4.170 & 0.532 $\pm\ 0.032^{+ 0.016 }_{- 0.001}$ & $\cdots$ \\
    4.180 & 0.714 $\pm\ 0.135^{+ 0.029 }_{- 0.001}$ & $\cdots$ \\
    4.200 & 0.895 $\pm\ 0.179^{+ 0.001 }_{- 0.057}$ & $\cdots$ \\
    4.260 & 0.703 $\pm\ 0.120^{+ 0.001 }_{- 0.090}$ & $0.393\ \pm\ 0.082 $ \\

\bottomrule
\end{tabular}%
\end{small}
\end{center}

\end{document}